\documentclass[aps,preprint,longbibliography,nofootinbib]{revtex4-1}

\usepackage{graphicx}
\usepackage{dcolumn}

\begin{document}

\title{Revisiting the $B^{(*)}_s$-Meson Production at the Hadronic Colliders}

\author{Jia-Wei Zhang}
\author{Xue-Wen Chen}
\affiliation{Department of Physics, Chongqing University of Science and Technology, Chongqing 401331, P.R. China}

\author{Jun Jiang}
\author{Zhan Sun}
\author{Xing-Gang Wu}
\email{wuxg@cqu.edu.cn}
\affiliation{Department of Physics, Chongqing University, Chongqing 401331, P.R. China}

\date{\today}

\begin{abstract}

The hadronic production of the heavy-flavored hadron provides a challenging opportunity to test the validity of pQCD predictions. In the paper, we make a comparative study on the properties of the $B^{(*)}_s$ hadroproduction within either the fixed-flavor-number scheme (FFNS) or the general-mass variable-flavor-number scheme (GM-VFNS). By using FFNS, as is previously adopted in the literature, one only needs to deal with the dominant gluon-gluon fusion mechanism via the subprocess $g+g\rightarrow B^{(*)}_s+b+\bar{s}$. While by using GM-VFNS, one needs to deal with two mechanisms: one is the gluon-gluon fusion mechanism and the other is the extrinsic heavy quark mechanism via the subprocesses $g+\bar{b}\to B^{(*)}_s +\bar{s}$ and $g+s\to B^{(*)}_s +b$. It is found that both mechanisms can provide reasonable contributions to the $B^{(*)}_s$ hadroproduction under the GM-VFNS, and there is double counting for those two mechanisms in specific kinematic regions. At the Tevatron, the differences between the estimations of FFNS and GM-VFNS are small, e.g. after cutting off the small $p_T$ events (cf. $p_T>4GeV$), the $B^{(*)}_s$ $p_T$-distributions are almost coincide with each other. However these differences are obvious at the LHC. The forthcoming more precise data on LHC shall provide a good chance to check which scheme is more appropriate to deal with the $B^{(*)}_s$-meson production and to further study the heavy quark components in hadrons.  \\

\begin{description}

\item[PACS numbers] 14.40.Nd, 14.40.Pq, 12.38.Bx

\end{description}

\end{abstract}

\maketitle

\section{Introduction}

The CDF and D0 Collaborations have successfully collected the $B$-hadron data since the Tevatron Run II started in 2001~\cite{CDF1,CDF2,D0}. The CERN LHC shall also provide a good platform to study the properties of $B$ hadron~\cite{LHC1,LHC2}. Recently, the CMS and ATLAS Collaborations at the LHC published first results for $B$-hadron production at $\sqrt{S}=7$ TeV~\cite{CMS1,CMS2,CMS3,ATLAS}. These measurements stimulate a significant improvement for the heavy-quark hadronproduction. In the present paper, we shall focus our attention on the $B^{(*)}_s$-meson hadroproduction.

At a hadronic collider, for the heavy meson/baryon production in higher transverse momentum and smaller rapidity regions, which corresponding to larger momentum fraction of the constitute quarks, it is found that the heavy quark components in proton or antiproton is always quite small in comparison to that of the light quarks or gluons. So, in the literature, one usually does not take the hadron's heavy quark components into consideration in most of the calculations for the heavy quarkonium production. Thus, the fixed-flavor-number scheme (FFNS) is usually adopted~\cite{ffn1,ffn2,ffn3}. Within FFNS, the number of active flavors in the initial hadron is fixed to be $n_f=3$, and then only light quarks/antiquarks and gluon should be considered in the initial state of the hard scattering subprocess. For example, we have studied the hadronic production of the spin-singlet $B_s$ and the spin-triplet $B^{*}_s$ by adopting FFNS~\cite{zfw}, in which the dominant gluon-gluon fusion mechanism via the subprocess $g+g\rightarrow B^{(*)}_s+b+\bar{s}$ has been studied.

It has been argued that in certain cases the heavy quark components in the collision hadrons may also provide sizable contributions in specific kinematic regions. For example, it has been shown that the mechanisms involving heavy quarks in the initial state can give sizable contributions to the hadronic production of $(c\bar{c})$-quarkonium~\cite{qiaocc}, $(c\bar{b})$-quarkonium~\cite{wubc}, $\Xi_{cc}$-baryon~\cite{wuxicc1,wuxicc2}, and etc.. Especially, it is found that in lower $p_T$ region, the mechanism from those heavy quarks in the initial state (hereafter refer to as ``heavy quark mechanism") can even dominant over other mechanisms.

The heavy flavored quarks in hadron can be generated through two different ways inside the incident hadrons. On the one hand, it can be perturbatively generated by gluon splitting, and hence, it is usually named as the `extrinsic' component. Thus, in addition to the gluon-gluon fusion, one also needs to deal with the extrinsic heavy quark mechanism via the subprocesses $g+\bar{b}\to B^{(*)}_s +\bar{s}$ and $g+s\to B^{(*)}_s +b$. It is noted that a full quantum chromodynamics (QCD) evolved heavy quark distribution functions, according to the Altarelli-Parisi equation, includes all the terms proportional to $\ln\left({\mu^2} /{m^2_Q}\right)$ ($Q$ being the heavy quark). Some of these terms also occur in the gluon-gluon fusion mechanism after doing the phase-space integration, so one has to deal with the double counting problem~\cite{cwz}. To solve such problem, we shall adopt the general-mass variable-flavor-number scheme (GM-VFNS)~\cite{gmvfn2,gmvfn3,gmvfn4}. In specific kinematic regions, the extrinsic mechanism can have sizable contribution, as a comprehensive analysis of the $B^{(*)}_s$ hadronproduction, it will be interesting to study the extrinsic heavy quarks' contributions and also to make a comparison of the results with those obtained from FFNS.

On the other hand, the heavy quarks can also be generated non-perturbatively and appears at or even below the energy scale of the heavy quark threshold, which can be named as `intrinsic' component~\cite{brodsky1,brodsky2}. The upper bound for the probability of intrinsic $c$-quark in hadron is about $1\%$~\cite{intri1,intri2}, which can lead to sizable effect for the $\Xi_{cc}$-baryon production~\cite{wuxicc1,wuxicc2}. However, for the present case of $B^{(*)}_s$ production, the intrinsic $b$-quark component in the hadron is quite small, since its probability is about one order lower than that of the intrinsic charm quark~\cite{intri3}, and it shall lead to negligible contribution to the production at the hadronic colliders, so we shall not consider it.

The paper is organized as follows. In Sec.II, we present the calculation technology for estimating the $B^{(*)}_s$ hadroproduction under the GM-VFNS. In Sec.III, we present our numerical results, a comparison of the hadronic production of $B^{(*)}_s$ from the GM-VFNS and the FFNS is also presented. The final section is reserved for a summary.

\section{Calculation Technology}

According to perturbative QCD factorization theorem, the cross-section for the hadronic production of $B^{(*)}_s$ under the GM-VFNS can be formulated as
\begin{widetext}
\begin{eqnarray}
d\sigma&=&F^{g}_{H_{1}}(x_{1},\mu) F^{g}_{H_{2}}(x_{2},\mu)
\bigotimes d\hat{\sigma}_{gg\rightarrow
B^{(*)}_s}(x_{1},x_{2},\mu)\nonumber\\
&+& \sum_{i,j=1,2;i\neq j}F^{g}_{H_{i}}(x_{1},\mu)
\left[F^{\bar{b}}_{H_{j}}(x_{2},\mu)-
F^{g}_{H_{j}}(x_{2},\mu)\bigotimes F^{\bar{b}}_g(x_2,\mu)\right]
\bigotimes d\hat{\sigma}_{g\bar{b}\rightarrow
B^{(*)}_s}(x_{1},x_{2},\mu)\nonumber\\
&+& \sum_{i,j=1,2;i\neq j}F^{g}_{H_{i}}(x_{1},\mu)
\left[F^{s}_{H_{j}}(x_{2},\mu)-
F^{g}_{H_{j}}(x_{2},\mu)\bigotimes F^{s}_g(x_2,\mu)\right]
\bigotimes d\hat{\sigma}_{gs\rightarrow
B^{(*)}_s}(x_{1},x_{2},\mu) + \cdots, \label{basic}
\end{eqnarray}
\end{widetext}
where the ellipsis stands for the terms in higher $\alpha_s$ order and those terms with quite small contributions, such as the light quark and light anti-quark collision mechanism and etc.. The first term in Eq.(\ref{basic}) shows the gluon-gluon fusion mechanism, which is dominant one under the FFNS. The second and third one represent the extrinsic $b$-quark and $s$-quark mechanism, in which the subtraction term is introduced to eliminate the double counting problem, respectively. In the present pQCD calculation, we treat $s$-quark as heavy quark in calculating the hard scattering amplitude. It is reasonable, since the effective $s$-quark mass for the pQCD calculation can be chosen around its constitute quark mass, whose standard value is about $490$ GeV within the constitute quark model~\cite{cqm1,cqm2}. Moreover, since the intermediate gluon should be hard enough so as to generate a heavy $(b\bar{b})$-pair or $(s\bar{s})$-pair, the typical energy scale $\mu \geq 2m_b (\;{\rm or}\;2m_s) > \Lambda_{QCD}$, these mechanisms are pQCD calculable. As shown by Ref.\cite{zfw}, under such treatment, we can obtain reasonable estimations for the total and differential cross sections.

The function $F^{i}_{H}(x,\mu)$ (with $H=H_1$ or $H_2$ and $x=x_1$ or $x_2$) stands for the distribution function of parton $i$ in hadron $H$. $d\sigma$ stands for the hadronic cross-section and $d\hat\sigma$ stands for the corresponding subprocesses. As a conventional treatment, we have taken the renormalization scale $\mu_R$ for the subprocess and the factorization scale $\mu_F$ for factorizing the parton distribution functions (PDFs) and the hard subprocess to be the same, i.e. $\mu_R=\mu_F=\mu$. And the subtraction term for $F^{Q}_{H}(x,\mu)$ is defined as
\begin{eqnarray}\label{subtraction}
F^{Q}_{H}(x,\mu)_{SUB}&=&F^{g}_{H}(x,\mu)\bigotimes
F^{Q}_g(x,\mu)= \int^1_{x}F^{Q}_g(\kappa,\mu)
F^{g}_{H}\left(\frac{x}{\kappa},\mu\right) \frac{d\kappa}{\kappa},
\end{eqnarray}
where the quark distribution $F^{Q}_g(x,\mu)$ (where $Q$ stand for heavy-quark $s$ or $\bar{b}$) is connected to the familiar $g\to Q\bar{Q}$ splitting function $P_{g\to Q}$, and its form can be written as
\begin{equation}\label{quark}
F^{Q}_g(x,\mu)=\frac{\alpha_s(\mu)}{4\pi}(1-2x+2x^2) \ln\frac{\mu^2}{m^2_{Q}}.
\end{equation}

\begin{figure}[htb]
\centering
\includegraphics[width=0.9\textwidth]{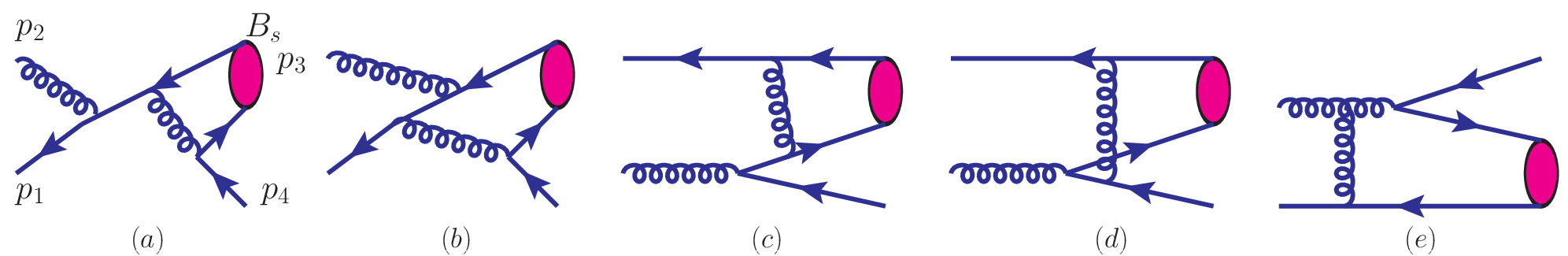}
\caption{Typical Feynman diagrams for the heavy quark mechanism, i.e. hadroproduction of $B^{(*)}_s$ via the subprocesses: $g(p_2)+ \bar{b}(p_1)\to B^{(*)}_s(p_3)+\bar{s}(p_4)$. The Feynman diagrams for the $g(p_2) +s(p_1)\to B^{(*)}_s(p_3) + b(p_4)$ can be obtained by the replacement $\bar{b}\to s$ and $\bar{s}\to b$. }
\label{feyn}
\end{figure}

As for the gluon-gluon fusion mechanism, we need to deal with 36 Feynman diagrams at the leading order $(\alpha_s^4)$~\cite{zfw}, where in different to the conventional squared amplitude approach, the improved helicity amplitude approach~\cite{bcvegpy1} has been adopted to get analytic and compact results as much as possible.

As for the extrinsic heavy quark mechanism for the hadronic $B^{(*)}_s$ production, we need to consider two subprocesses $g(p_2)+ \bar{b}(p_1)\to B^{(*)}_s(p_3)+\bar{s}(p_4)$ and $g(p_2) +s(p_1)\to B^{(*)}_s(p_3) + b(p_4)$, whose typical Feynman diagrams for the extrinsic bottom/strange mechanisms at the leading order (LO) are shown in Fig.(\ref{feyn}). In Eq.(\ref{basic}), the $d\hat{\sigma}_{ij\rightarrow B^{(*)}_s} (x_{1},x_{2},\mu)$ stands for the usual 2-to-2 differential cross section,
\begin{widetext}
\begin{equation}
d\hat{\sigma}_{ij\rightarrow B^{(*)}_s}(x_{1},x_{2},\mu) =\frac{(2\pi)^4|\overline{M}|^2} {4\sqrt{(p_1\cdot p_2)^2-p_{1}^{2} p_{2}^{2}}} \prod_{i=3}^{4}\frac{d^3\mathbf{p}_i} {(2\pi)^3(2E_i)}\delta \left(\sum_{i=3}^{4}p_i-p_1-p_2\right) \;,
\end{equation}
\end{widetext}
where $i\neq j$ and $i,j=g,\bar{b}$ for the extrinsic $b$-quark mechanism and $i,j=g,s$ for the extrinsic $s$-quark mechanism and the initial-parton spin and color average and the final-state quantum number summation are all attributed to $|\overline{M}|^2$. $|\overline{M}|^2$ can be calculated by using the conventional squared-amplitude approach, only one needs to keep the heavy quark mass terms according to the GM-VFNS. For shortening the text, we do not put $|\overline{M}|^2$ here, whose explicit form can be found in the appendix of Ref.\cite{wubc} by simply changing $m_c$ there to be the present case of $m_s$. Next, to accomplish the phase space integration, we adopt two subroutines RAMBOS~\cite{rkw} and VEGAS~\cite{gpl}, which together with some reasonable transformations to make them run more effectively can be found in the generators BCVEGPY~\cite{bcvegpy2} and GENXICC~\cite{genxicc1}.

\begin{figure}
\centering
\includegraphics[width=0.8\textwidth]{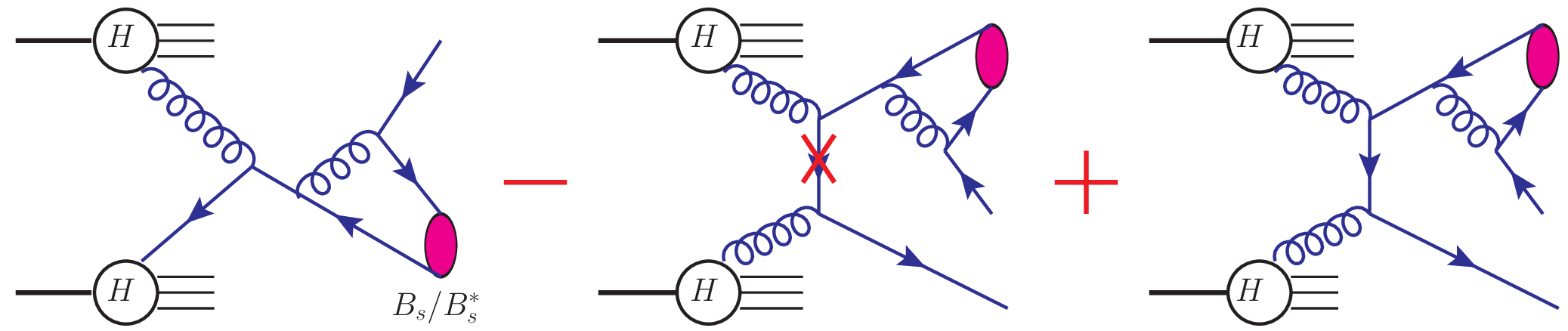}
\caption{Graphical representation for the subtraction method within the GM-VFNS. The symbol $\times$ on the internal quark line in the subtraction term indicates that it is close to the mass-shell and collinear to the gluon and hadron momentum. The combination of the first and the second terms are called as extrinsic mechanism. } \label{graphical} \vspace{-0mm}
\end{figure}

As has been discussed in the Introduction, to deal with the $B^{(*)}_s$ production through the heavy quark mechanism and the gluon-gluon fusion mechanism simultaneously, one needs to solve the double counting problem. This is due to fact that in some of the Feynman diagrams of the gluon-gluon fusion mechanism, when the intermediate heavy quark line that is next to the incident gluon is nearly on shell and is collinear to the incident gluon, then it will result in a factor of order $\alpha_s$ distribution of a quark in a gluon, like Eq.(\ref{quark}). More explicitly, we draw Fig.(\ref{graphical}) as an example to graphically illustrate this point. In Fig.(\ref{graphical}), the symbol $(\times)$ on the internal quark line in the subtraction term means that the heavy quark is on mass-shell and moving longitudinally. The strict GM-VFNS needs a full NLO calculation~\cite{gmvfn1,gmvfn2,gmvfn3,gmvfn4}, which is not available at the present due to its complexity. As a try to solve the double counting problem, we adopt the simplified version of GM-VFN to do our analysis, i.e. the Aivazis-Collins-Olness-Tung (ACOT) scheme~\cite{gmvfn1,acot}, in which, only the dominant leading log terms are taken into consideration and are absorbed into the redefinition of the parton distribution functions.

\section{Numerical Results and Discussions}

In doing numerical calculation, we adopt the Lattice QCD result for $f_{B_{s}}$, i.e. $f_{B_{s}} =0.232$ GeV~\cite{lattice,potential}. The masses of $b$ and $s$ quarks are taken as $m_b=4.90 {\rm GeV}$ and $m_s=0.50 {\rm GeV}$, and to ensure the gauge invariance of the hard scattering amplitude, the mass of the bound state is taken to be the sum of the two constitute quark masses, i.e. $M_{B_{s}}=m_b+m_s$. Because the spin splitting effect is ignorable here, there is no difference for the decay constant and the mass between the spin states $B_{s}[^1S_0]$ and $B_{s}^{*}[^3S_1]$. The scale $\mu$ is set to be the transverse mass of the bound state, i.e. $\mu=\sqrt{M^2_{B^{(*)}_s}+p_{T}^2}$, where $p_T$ is the transverse momentum of the bound state. When using the GM-VFNS, CTEQ6HQ~\cite{6hqcteq} is adopted for PDF, and to be consistent, the NLO $\alpha_s$ running above $\Lambda^{(n_f=4)}_{QCD}=0.326$ GeV is adopted, i.e. $\alpha_s(\mu^2)=\frac{4\pi}{\beta_0\ln(\mu^2/\Lambda^2_{QCD})} \left[1-\frac{2\beta_1} {\beta_0^2} \frac{\ln[\ln(\mu^2/\Lambda_{QCD}^2)]}{\ln(\mu^2/\Lambda_{QCD}^2)}\right]$, where $\beta_0=11-2n_f/3$ and $\beta_1=51-19n_f/3$. It is noted that for the GM-VFNS, the active flavor number $n_f$ changes with the energy scale. As a comparison, the LO $\alpha_s$ running with fixed $n_f=3$ and the typical PDF CTEQ6L1~\cite{6lcteq} are adopted for the FFNS.

\subsection{$B^{(*)}_s$ Hadroproduction from the extrinsic mechanisms }

\begin{figure}[ht]
\centering
\includegraphics[width=0.6\textwidth]{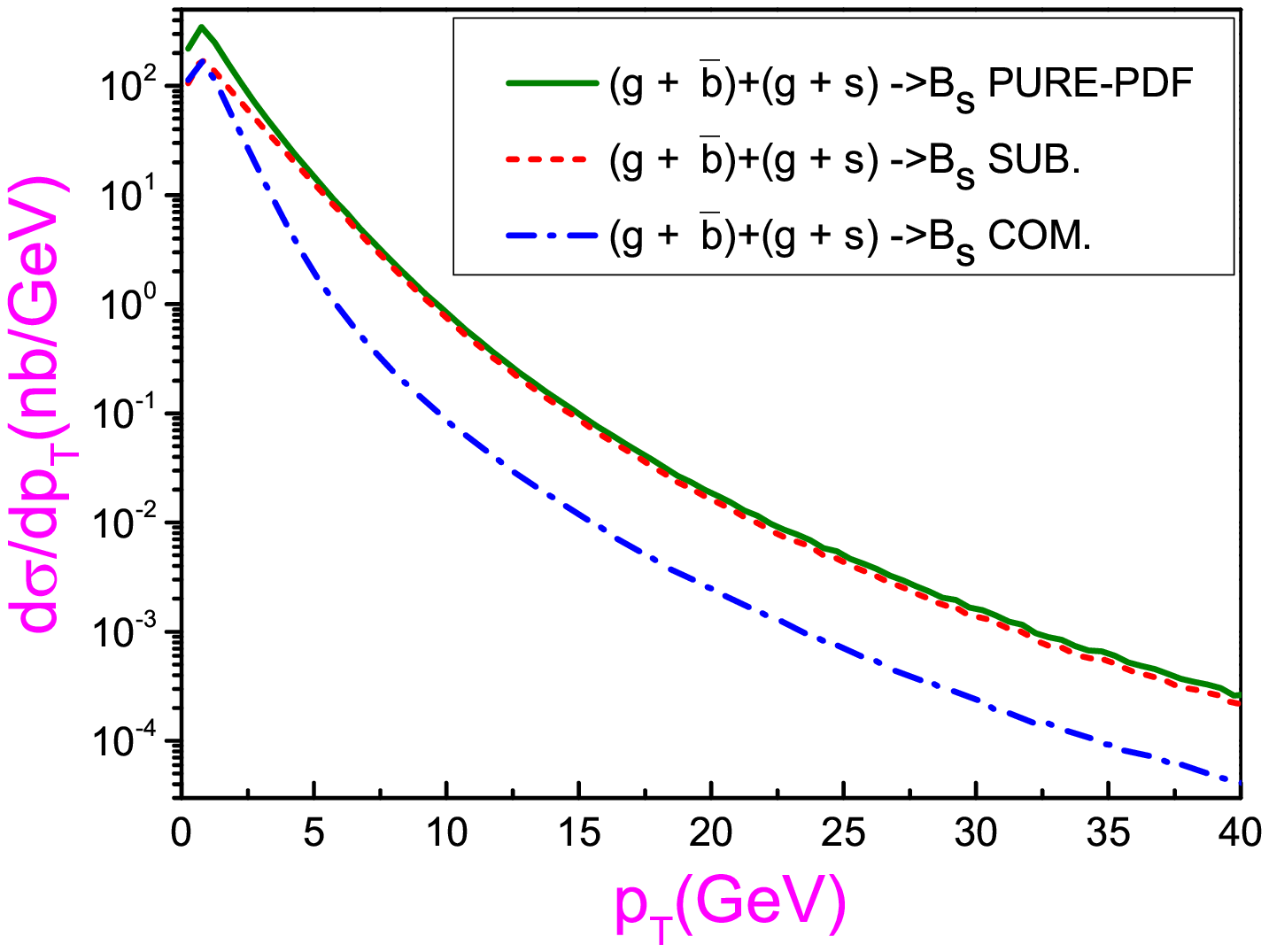}
\caption{$p_T$-distributions for the hadronic production of $B_s$ at LHC with $\sqrt S=14.0$ TeV, where $(g+\bar{b})+(g+s)$ represents the sum of the extrinsic $b$-quark and $s$-quark mechanisms. The `PURE-PDF' means the heavy quarks' PDF are taken as CTEQ6HQ, the `SUB.' means the heavy quarks' PDF are taken to be the subtraction term defined in Eq.(\ref{subtraction}), the `COM.' stands for the combination of the `PURE-PDF' and `SUB.' components as indicated by Eq.(\ref{basic}). } \label{sub}
\end{figure}

Firstly, it is interesting to show how the double counting term is subtracted under GM-VFNS. We present the results for the extrinsic $b$-quark and $s$-quark mechanisms for the hadroproduction of $B_s$ at LHC in Fig.(\ref{sub}). It can be found that there exists a large cancelation between the contributions from the `PURE-PDF' term (with the heavy quarks' PDF taken to be CTEQ6HQ) and the subtraction term (with the heavy quarks' PDF taken to be the subtraction term defined in Eq.(\ref{subtraction})) \footnote{Such large cancelation is reasonable, since the `SUB.'-term as defined by Eq.(\ref{subtraction}) provides the leading log contribution to the heavy quark PDF. At small $p_T$ region, most of the events are small $x$ events, and the differences for the cross-sections are further amplified by large values of PDFs at small $x$ region. This conceptually explains why there is large cancelation at high $p_T$ but not at small $p_T$ regions.}. When taking the contribution of both the extrinsic $b$-quark and $s$-quark mechanisms into account, a large cancelation can be found in large $p_T$ regions. This shows clearly that to obtain a reliable result, we should take such double counting term into consideration, otherwise, the result with be highly overestimated.

\begin{table}[ht]
\begin{center}
\caption{Cross section (in unit $nb$) for the hadronic production of $B_s$ at LHC I with $\sqrt S=8.0$ TeV and LHC II with $\sqrt S=14.0$ TeV and Tevatron with $\sqrt{S}=1.96$ TeV under GM-VFNS and FFNS, where the $(g+\bar{b})+(g+s)$ represents the sum of the extrinsic $b$-quark and $s$-quark mechanisms and etc. Three typical $p_T$ cuts are adopted. As for the rapidity cut, we take $|y|\leq 1.5$ for LHC and $|y|\leq 0.6$ for Tevatron. } \vspace{2mm}
\begin{tabular}{|c|c|ccc|c|}
\hline
-&-& \multicolumn{3}{c|}{GM-VFNS}& FFNS \\
$\sqrt S$&$p_{Tcut}$&$(g+\bar{b})+(g+s)$ & $g+g$ & {\it total}&  $g+g$ \\
\hline
LHC I  &$0$ GeV   &80.38 & 51.78 &132.2 &82.87 \\
-      &$2.5$ GeV &8.30 & 36.70 &45.00 &57.90 \\
-      &$4.0$ GeV &2.13 & 24.19 &26.32 &37.09 \\
\hline
LHC II &$0$ GeV   &125.7 & 78.46 &204.1 &138.7 \\
-      &$2.5$ GeV &13.20 & 56.85 &70.05 &98.28 \\
-      &$4.0$ GeV &3.44 & 38.28 &41.72 &64.01 \\
\hline
Tevatron
       &$0$ GeV &11.27 & 6.75 &18.01&8.35 \\
-      &$2.5$ GeV &1.27 & 4.41 &5.68&5.45 \\
 -     &$4.0$ GeV &0.32 & 2.73 &3.05&3.30 \\
 \hline
\end{tabular}\label{cutcross1}
\end{center}
\end{table}

\begin{table}[ht]
\begin{center}
\caption{Cross section (in unit $nb$) for the hadronic production of $B_s^*$ at LHC I with $\sqrt S=8.0$ TeV and LHC II with $\sqrt S=14.0$ TeV and Tevatron with $\sqrt{S}=1.96$ TeV under GM-VFNS and FFNS, where the $(g+\bar{b})+(g+s)$ represents the sum of the extrinsic $b$-quark and $s$-quark mechanisms and etc. Three typical $p_T$ cuts are adopted. As for the rapidity cut, we take $|y|\leq 1.5$ for LHC and $|y|\leq 0.6$ for Tevatron.} \vspace{2mm}
\begin{tabular}{|c|c|ccc|c|}
\hline
-&-& \multicolumn{3}{c|}{GM-VFNS}& FFNS \\
$\sqrt S$&$p_{Tcut}$&$(g+\bar{b})+(g+s)$ & $g+g$ & {\it total}&  $g+g$ \\
\hline
LHC I       &$0$ GeV   &141.1 &169.8  &310.9 &274.0 \\
-          &$2.5$ GeV &15.89 & 115.9 &131.8 &184.2 \\
-          &$4.0$ GeV &2.88 & 73.87 &76.75 &113.9 \\
\hline
LHC II      &$0$ GeV   &224.1& 258.0 &482.1 & 460.9 \\
-          &$2.5$ GeV &25.22& 180.6 &205.8 & 316.3 \\
-          &$4.0$ GeV &4.45& 117.4 &121.9 & 199.6 \\
\hline
Tevatron
             &$0$ GeV   &15.69& 22.63 &38.32 &28.02 \\
-          &$2.5$ GeV &2.40&14.13  &16.53 &17.55\\
-         &$4.0$ GeV &0.53&8.49 &9.02 & 10.29\\
\hline
\end{tabular}
\label{cutcross2}
\end{center}
\end{table}

Secondly, we present the total cross sections for the hadroproduction of $B_s$ and $B^{*}_s$ at LHC I with $\sqrt S=8.0$ TeV and LHC II with $\sqrt S=14.0$ TeV, and Tevatron with $\sqrt S=1.96$ TeV, which are presented in Tables \ref{cutcross1} and \ref{cutcross2}. Three typical transverse momentum cuts $p_{Tcut}=0$ GeV, $p_{Tcut}=2.5$ GeV and $p_{Tcut}=4$ GeV for both LHC and Tevatron, and rapidity cut $|y|\leq 1.5$ for LHC, $|y|\leq 0.6$ for Tevatron are adopted in the calculation. Tables \ref{cutcross1} and \ref{cutcross2} show that the total cross sections of the extrinsic $b$-quark and $s$-quark mechanisms are comparable to those of the gluon-gluon fusion mechanism under GM-VFNS. The large cross sections of the extrinsic $b$-quark and $s$-quark mechanisms mainly come from small $p_T$ region. For example, when $p_{Tcut}=0$ GeV, the ratio between the total cross-section of the extrinsic $b$-quark and $s$-quark mechanisms and that of the gluon-gluon fusion mechanism are $160\%$ for $B_s$ and $80\%$ for $B^{*}_s$ at both LHC and Tevatron; and when $p_{Tcut}$ increases to 2.5 GeV and 4.0 GeV, such ratio changes down to $\sim 23\%$ and $\sim 9\%$ for the case of $B_s$, and $\sim 15\%$ and $\sim 5\%$ for the case of $B^{*}_s$, respectively.

\begin{figure}[htb]
\centering
\includegraphics[width=0.49\textwidth]{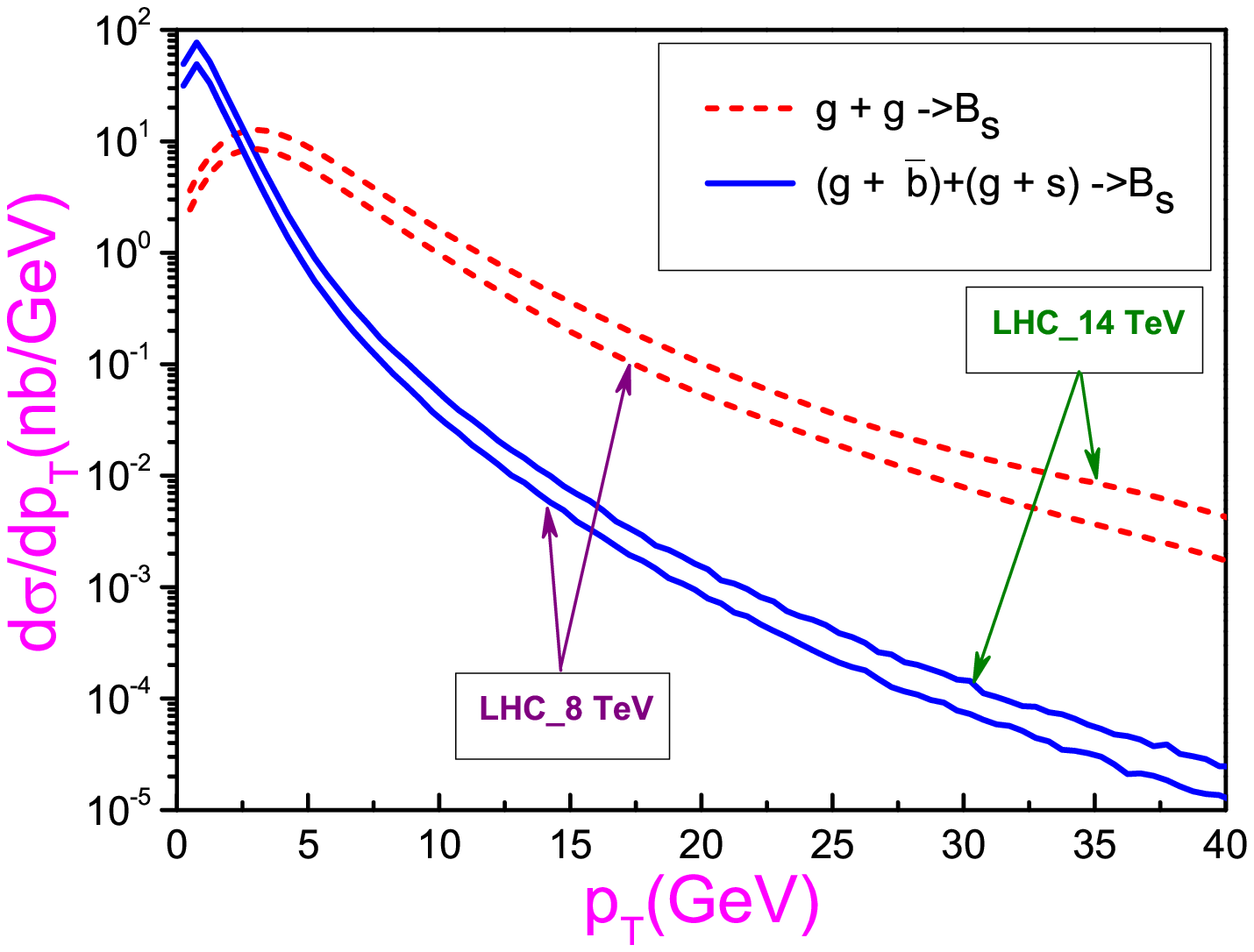}
\includegraphics[width=0.49\textwidth]{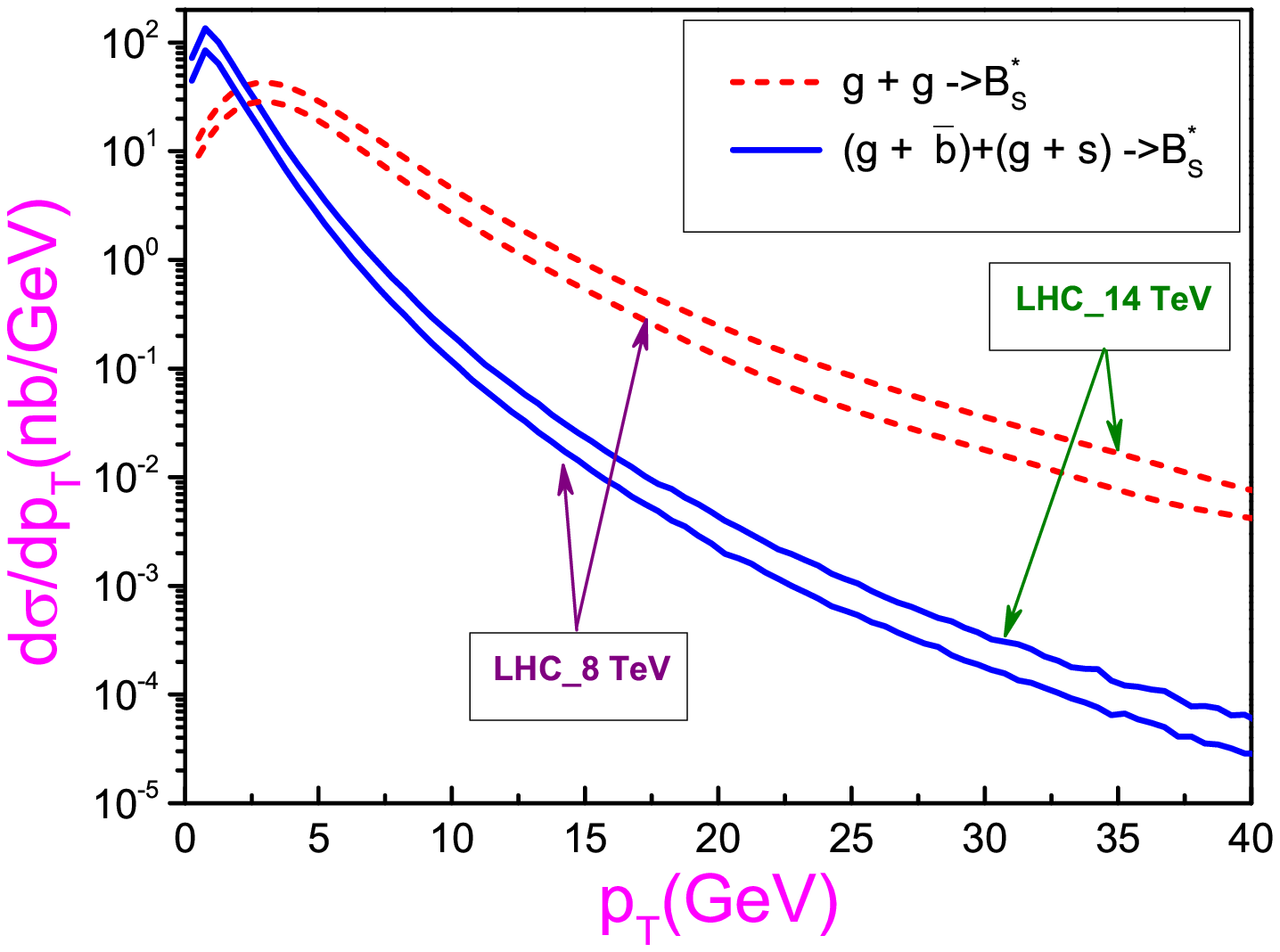}
\caption{$p_T$-distributions for the hadronic production of $B_s$ (Upper) and $B_s^*$ (Lower) at LHC with $\sqrt S=8.0$ TeV and $\sqrt S=14.0$ TeV under the GM-VFNS. The dashed and the solid lines represent the gluon-gluon fusion mechanism and the extrinsic heavy quark mechanism, respectively, where the $(g+\bar{b})+(g+s)$ represents the sum of the extrinsic $b$-quark and $s$-quark mechanisms. All $p_T$ distributions are drawn under $|y|<1.5$ and the PDF is taken as CTEQ6HQ.} \label{fig1}
\end{figure}

\begin{figure}[htb]
\centering
\includegraphics[width=0.49\textwidth]{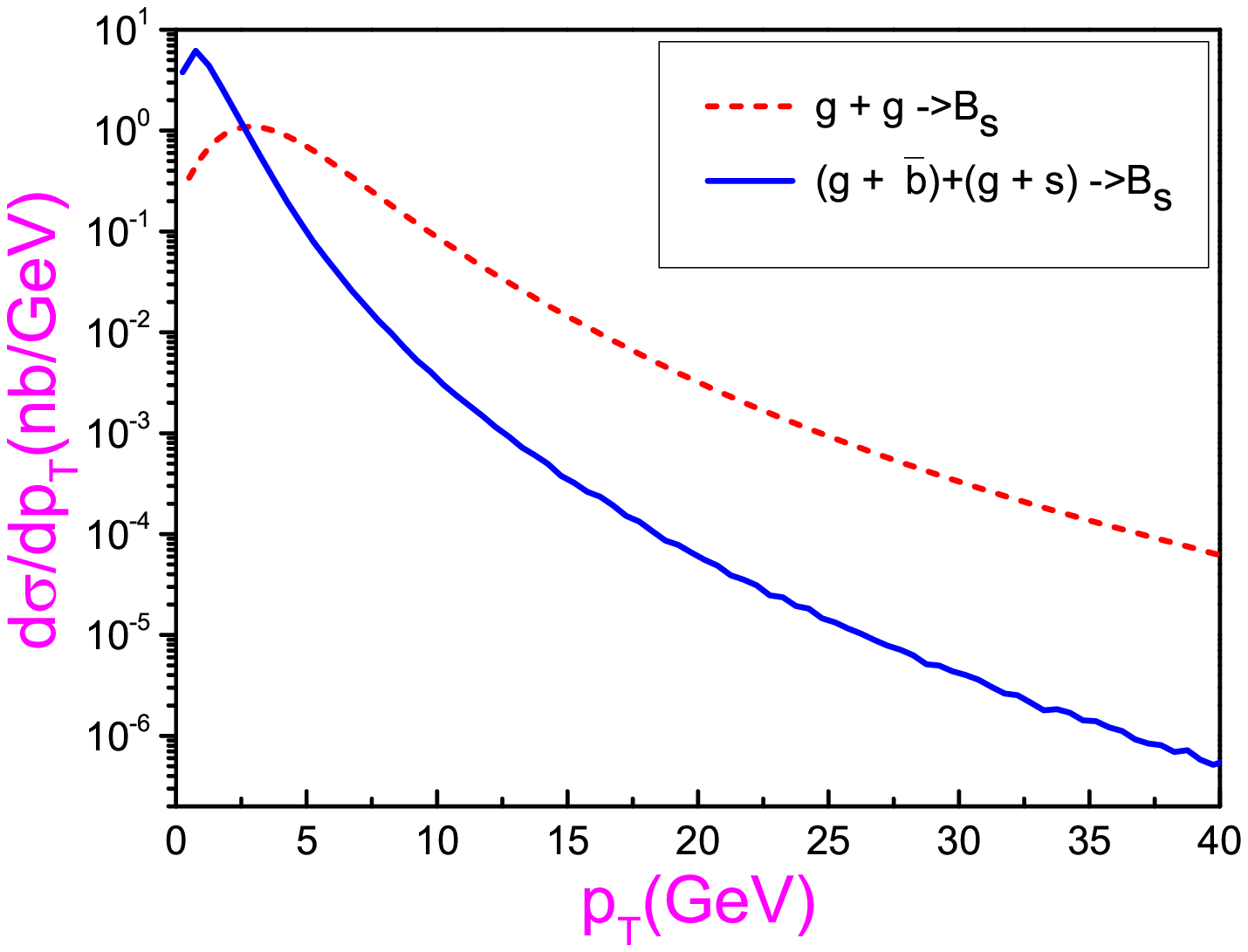}
\includegraphics[width=0.49\textwidth]{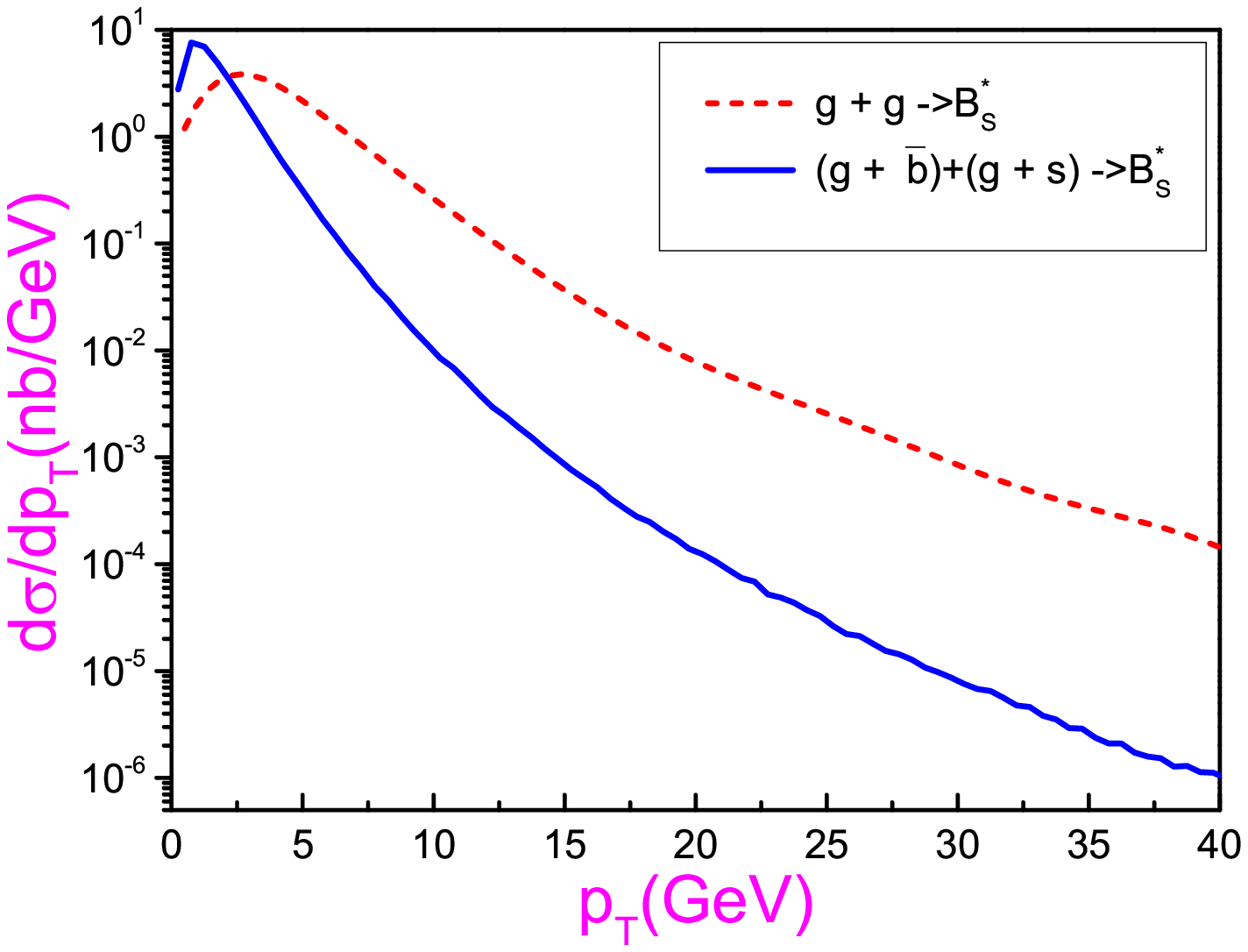}
\caption{$p_T$ distributions for the hadronic production of $B_s$ (Upper) and $B_s^*$ (Lower) at Tevatron under the GM-VFNS. The dashed and the solid lines represent the gluon-gluon fusion mechanism and the extrinsic heavy quark mechanism, respectively, where the $(g+\bar{b})+(g+s)$ represents the sum of the extrinsic $b$-quark and $s$-quark mechanisms and etc. All $p_T$ distributions are drawn under $|y|<0.6$ and the PDF is taken as CTEQ6HQ.} \label{fig2}
\end{figure}

Moreover, in order to illustrate this point clearly, we draw the transverse momentum $p_T$ distributions of $B^{(*)}_s$ in Fig.(\ref{fig1}) for LHC with $\sqrt S=8.0$ TeV and $\sqrt S=14.0$ TeV, and Fig.(\ref{fig2}) for Tevatron with $\sqrt{S}=1.96$ TeV. For the extrinsic heavy quark mechanisms, they are $2\to2$ subprocesses, and it is reasonable that the dominant distributions are in small $p_T$ regions. These figures show obviously that in small $p_T$ region, the contribution of the extrinsic $b$-quark and $s$-quark mechanisms is greater than that of the gluon-gluon fusion mechanism. However, it drops down quickly with the increment of $p_T$, and in large $p_T$ region, the contribution from the extrinsic strange and bottom mechanisms will be highly suppressed than that of the gluon-gluon fusion mechanism.

\subsection{A simple discussion on the uncertainties from the $b$-quark mass}

The uncertainties for the hadronic production of $B_s$ and $B^{*}_s$ include the PDFs, the quark masses, the factorization scale and etc.. Here, we will concentrate our attention on the $b$-quark mass effect \footnote{Other uncertainty sources shall give similar behaviors under both GM-VFNS and FFNS, which has been deeply analyzed under FFNS in Ref.\cite{zfw}, and to short the paper, we do not present extra discussions on other uncertainties.}.

\begin{table}
\begin{center}
\caption{Cross-section (in unit $nb$) for the hadronic production of $B_s[1^{1}S_{0}]$ at LHC I with $\sqrt S=8.0$ TeV and LHC II with $\sqrt S=14.0$ TeV and Tevatron with $\sqrt{S}=1.96$ TeV under GM-VFNS and FFNS with $m_b\in[4.8,5.0]$ GeV, where the [$(g+\bar{b})+(g+s)$] represents the sum of the extrinsic $b$-quark and $s$-quark mechanism and etc. As for the rapidity cut, we take $|y|\leq 1.5$ for LHC and $|y|\leq 0.6$ for Tevatron. The upper, the center and the lower values are for $m=5.0$, $4.9$ and $4.8$ GeV respectively. } \vspace{2mm}
\begin{tabular}{|c|c|ccc|c|}
\hline
-&-& \multicolumn{3}{c|}{GM-VFNS}& FFNS \\
\hline $\sqrt S$ & $p_{Tcut}$ & [($g+\bar{b}$)+($g+s$)]& $g+g$ & {\it total}&  $g+g$\\
\hline LHC I &$2.5$  &$8.30^{+1.18}_{-1.31}$ & $36.70^{-1.19}_{+1.06}$ &$45.00^{-0.01}_{-0.25}$ & $57.90^{-1.71}_{+1.96}$\\
&$4.0$  & $2.13^{+0.31}_{-0.32}$ & $24.19^{-0.49}_{+0.40}$ & $26.32^{-0.18}_{+0.08}$ & $37.09^{-0.67}_{+0.72}$\\
\hline
 LHC II&$2.5$  &$13.20^{+1.81}_{-2.00}$ & $56.85^{-1.58}_{+1.78}$ &$70.05^{+0.23}_{-0.22}$ & $98.28^{-2.29}_{+4.12}$\\
&$4.0$  & $3.44^{+0.48}_{-0.50}$ & $38.28^{-0.66}_{+0.73}$ & $41.72^{-0.18}_{+0.23}$ & $64.01^{-0.64}_{+2.03}$\\
\hline
Tevatron &$2.5$  &$1.27^{+0.17}_{-0.20}$ & $4.41^{-0.16}_{+0.18}$ &$5.68^{-0.01}_{-0.02}$ & $5.45^{-0.22}_{+0.24}$\\
&$4.0$ & $0.32^{+0.04}_{-0.05}$ & $2.73^{-0.07}_{+0.07}$ &$3.05^{-0.03}_{+0.02}$ & $3.30^{-0.09}_{+0.10}$\\
\hline
\end{tabular}\label{mbptcutI}
\end{center}
\end{table}

\begin{table}[ht]
\begin{center}
\caption{Cross-section (in unit $nb$) for the hadronic production of $B^{*}_s[1^{3}S_{1}]$ at LHC I with $\sqrt S=8.0$ TeV and LHC II with $\sqrt S=14.0$ TeV and Tevatron with $\sqrt{S}=1.96$ TeV under GM-VFNS and FFNS with $m_b\in[4.8,5.0]$ GeV, where the [$(g+\bar{b})+(g+s)$] represents the sum of the extrinsic $b$-quark and $s$-quark mechanism and etc. As for the rapidity cut, we take $|y|\leq 1.5$ for LHC and $|y|\leq 0.6$ for Tevatron. The upper, the center and the lower values are for $m=5.0$, $4.9$ and $4.8$ GeV respectively. } \vspace{2mm}
\begin{tabular}{|c|c|ccc|c|}
\hline
-&-& \multicolumn{3}{c|}{GM-VFNS}& FFNS \\
\hline $\sqrt S$ & $p_{Tcut}$ & [($g+\bar{b}$)+($g+s$)]& $g+g$ & {\it total}&  $g+g$\\
\hline LHC I &$2.5$  &$15.89^{+2.75}_{-3.08}$ & $115.9^{-2.9}_{+4.0}$ &$131.8^{-0.2}_{+0.9}$ & $184.2^{-5.5}_{+6.0}$\\
&$4.0$  & $2.88^{+0.84}_{-0.89}$ & $73.87^{-0.88}_{+1.53}$ & $76.75^{-0.04}_{+0.64}$ & $113.9^{-2.1}_{+2.0}$ \\
\hline
 LHC II &$2.5$  &$25.22^{+4.22}_{-4.76}$ & $180.6^{-4.6}_{+4.0}$ &$205.8^{-0.38}_{-0.76}$ & $316.3^{-9.7}_{+9.5}$\\
&$4.0$  & $4.45^{+1.33}_{-1.41}$ & $117.4^{-1.3}_{+1.2}$ & $121.9^{-0.0}_{+0.2}$ & $199.3^{-3.1}_{+3.5}$\\
\hline Tevatron &$2.5$  &$2.40^{+0.38}_{-0.43}$ & $14.13^{-0.53}_{+0.52}$ &$16.53^{-0.15}_{+0.09}$ & $17.55^{-0.66}_{+0.68}$\\
&$4.0$ & $0.53^{+0.11}_{-0.12}$ & $8.49^{-0.20}_{+0.16}$ &$9.02^{-0.09}_{+0.05}$ & $10.29^{-0.26}_{+0.23}$\\
\hline
\end{tabular}\label{mbptcutII}
\end{center}
\end{table}

For the purpose of discussing the uncertainties caused by $m_b$, we vary $m_b\in[4.8,5.0]$ GeV, while other parameters are fixed to be their central values. The total cross section for the hadronic production of the scalar $B_s[1^{1}S_{0}]$ and the vector $B_s^*[1^{3}S_{1}]$ at LHC and Tevatron under GM-VFNS and FFNS are presented in Tables \ref{mbptcutI} and \ref{mbptcutII}.

From Tables \ref{mbptcutI} and \ref{mbptcutII}, it is found that the total cross section for the extrinsic strange and bottom mechanisms {\it increases} with the increment of $m_b$. For examples, when $m_b$ is increased by $0.1$ GeV, setting $p_{Tcut}=2.5$ GeV, the total cross sections for $B_s$ and $B_s^*$ will be {\it increased} by $16\%$ and $20\%$ at both LHC and Tevatron; while setting $p_{Tcut}=4.0$ GeV, such ratio changes to $15\%$ for $B_s$ and $30\%$ for $B_s^*$ at both LHC and Tevatron. Inversely, since the allowed phase space becomes narrower with the increment of $m_b$, it is found that the total cross section for the gluon-gluon fusion mechanism {\it decreases} with the increment of $m_b$. The total cross section of $B_s$ for the gluon-gluon fusion mechanism shall be {\it decreased} by $3\%$ ($4\%$) under both GM-VFNS and FFNS when $m_b$ increased by $0.1$ GeV for $p_{Tcut}\geq 2.5$ GeV at LHC (Tevatron). As a combination, due to the different behavior of the extrinsic mechanism and the gluon-gluon fusion mechanism, the total cross sections under GM-VFNS possess smaller uncertainties in comparison to that of FFNS, which is $\sim 0.5\%$ and $\sim 1\%$ at $p_{Tcut}=2.5$ GeV and $p_{Tcut}=4.0$ GeV, respectively, when $m_b$ increased by $0.1$ GeV at both LHC and Tevatron. This in some sense shows that the GM-VFNS treatment is more viable and leads to a more steady estimation.

\subsection{A comparison of GM-VFNS and FFNS}

We make a discussion on $B_s^{(*)}$ hadroproduction under GM-VFNS and FFNS. As for FFNS, we take PDF to be CTEQ6L1 \cite{6lcteq} and $\alpha_s$ to be at leading order. It should be noted that $n_f$ should be fixed to be $3$ in the FFNS and then to be consistent with the exact FFNS, the PDFs for the initial partons should be taken the one like CTEQ5F3 \cite{cteq5}, which is generated by using the evolution kernels with effective flavor number $n_{eff}=3$. As argued in Refs.\cite{changwu1,changwu2}, the uncertainties from different LO PDFs are small, and our numerically calculation shows that it only gives several percent difference by replacing CTEQ6L1 to CTEQ5F3 \footnote{Under FFNS, by varying the flavor number with the energy scale, the value of $\alpha_s$ shall be decreased, but this is to a large degree compensated by a larger gluon distribution function (i.e. in small $x$-region that is dominant for the production, $F^g_H({\rm CTEQ6L1})>F^g_H({\rm CTEQ5F3})$), so as a whole, there is small difference by using CTEQ6L1 and CTEQ5F3.}, so as a conventional choice, we adopt CTEQ6L1 as the typical PDF for FFNS.

Because the gluon distribution of CTEQ6HQ is always smaller than that of CTEQ6L1, especially in small $x$ region, so total cross section for gluon-gluon fusion mechanism under GM-VFNS is smaller than that under FFNS. Furthermore, since $x$ may reach up to much smaller region at LHC than at Tevatron, the difference between these two schemes is bigger at LHC than that at Tevatron. Tables \ref{cutcross1} and \ref{cutcross2} show this point clearly. For example, for the hadroproduction of $B_s$, when $p_{Tcut}=2.5$ GeV, it is found that at LHC, the total cross section for the gluon-gluon fusion under the GM-VFNS is only $58\%$ of that of FFNS; while at Tevatron, such ratio raises up to $81\%$. When $p_{Tcut}=4.0$ GeV, the change of ratios is very tiny. This shows that when taking the extrinsic mechanisms into account for the GM-VFNS, one can shrink the gap between the GM-VFNS and the FFNS results to a certain degree.

\begin{figure}[ht]
\centering
\includegraphics[width=0.49\textwidth]{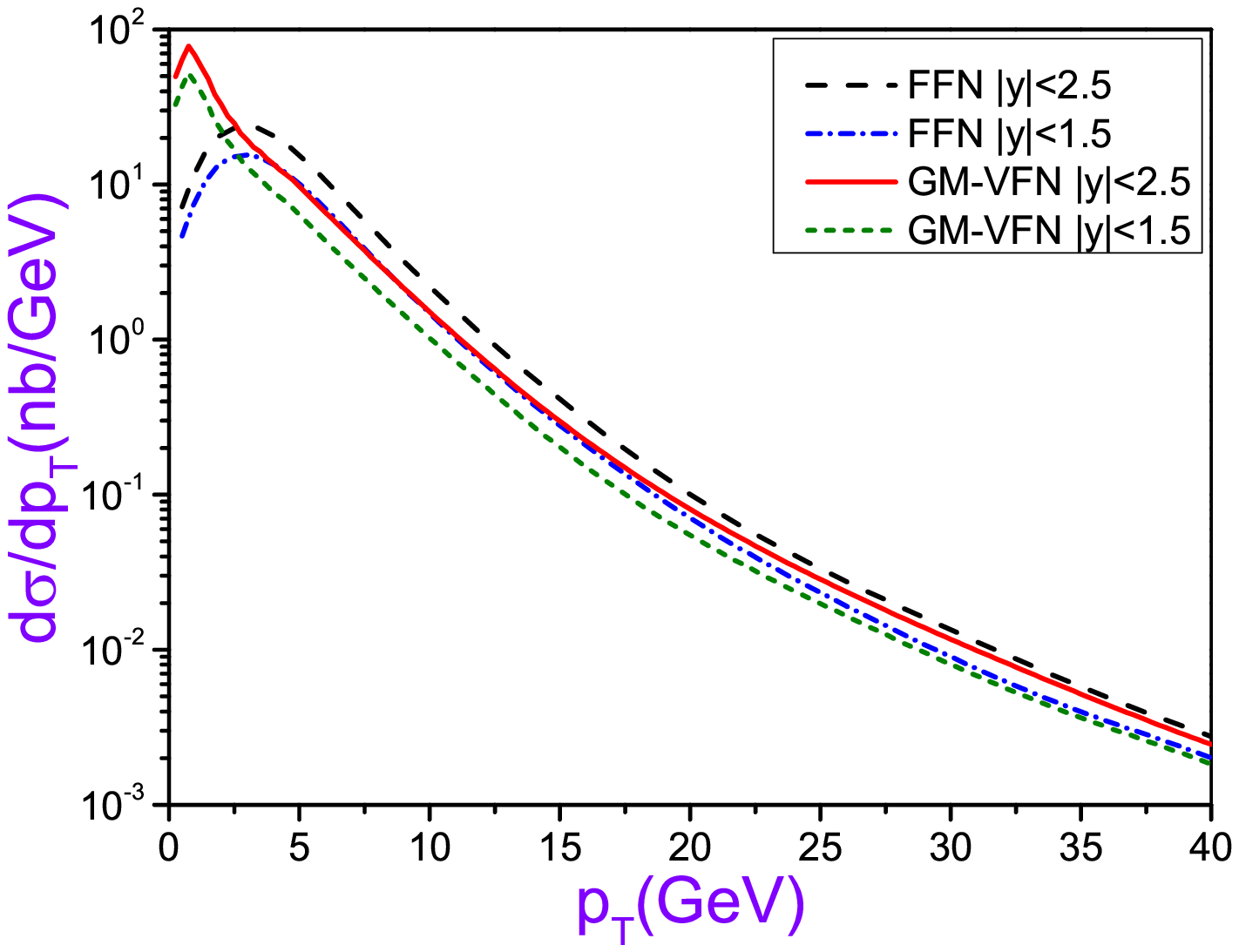}
\includegraphics[width=0.49\textwidth]{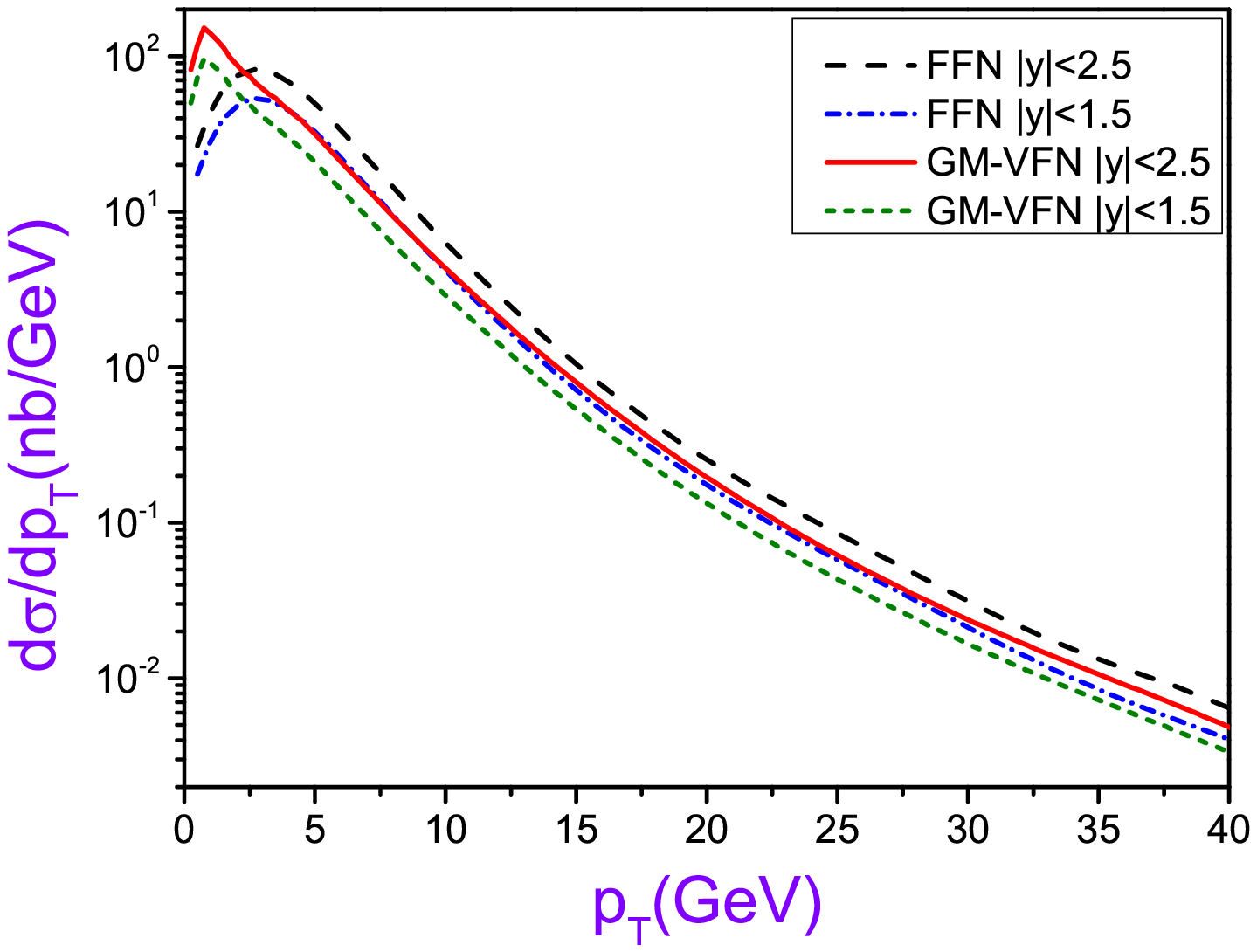}
\caption{$p_T$ distributions for the hadronic production of $B_s$ (Left) and $B_s^*$ (Right) at LHC with $\sqrt S=8.0$ TeV. The dashed and the dash-doted lines are for gluon-gluon fusion results obtained under the FFNS for rapidity cuts $|y|<2.5$ and $|y|<1.5$, respectively. The solid and the short-dashed lines stand for the total (The sum of the extrinsic $b$-quark and $s$-quark mechanism and the gluon-gluon fusion mechanism) results obtained under the GM-VFNS for rapidity cuts $|y|<2.5$ and $|y|<1.5$ respectively.} \label{fig3}
\end{figure}

\begin{figure}[ht]
\centering
\includegraphics[width=0.49\textwidth]{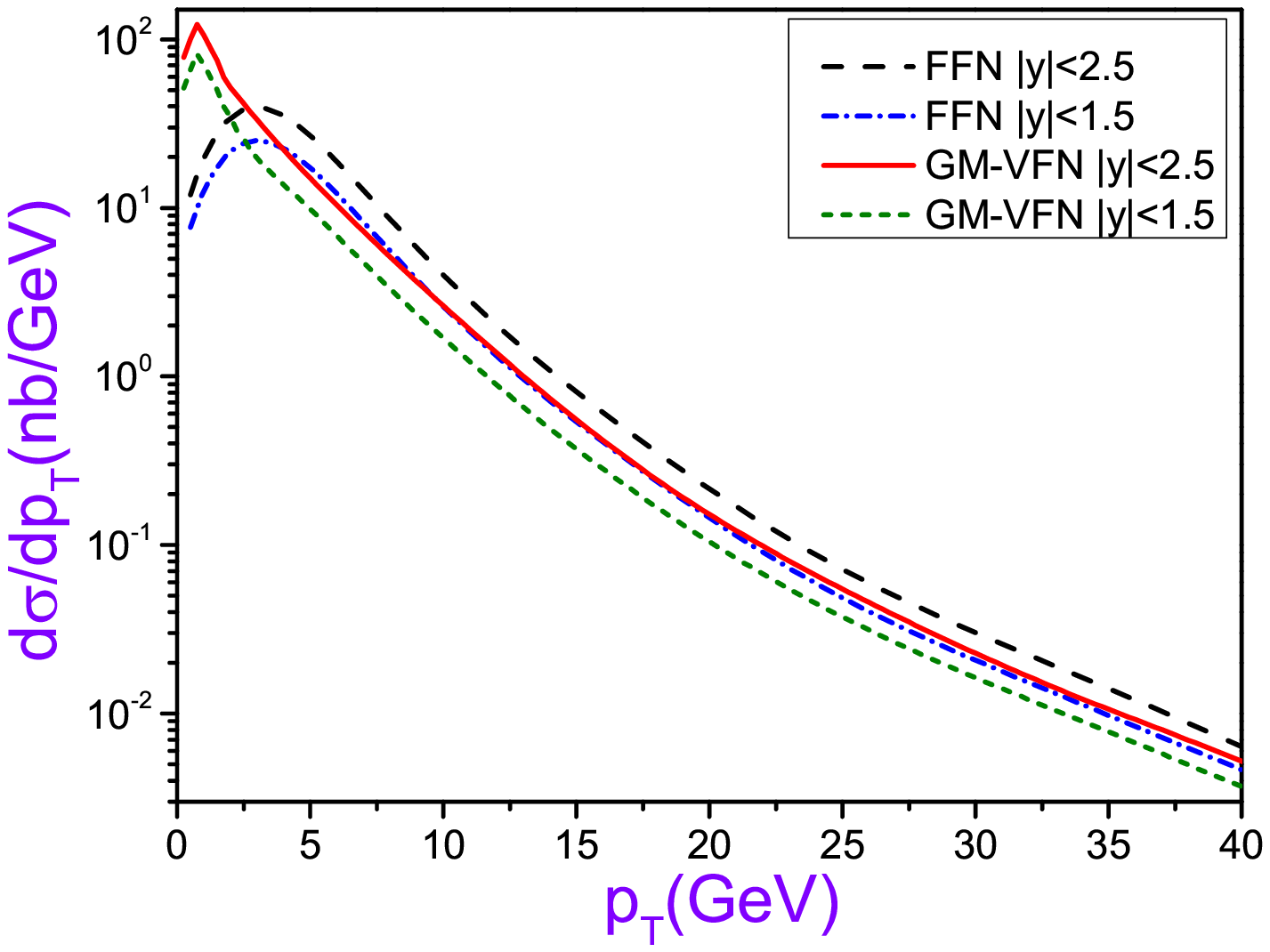}
\includegraphics[width=0.49\textwidth]{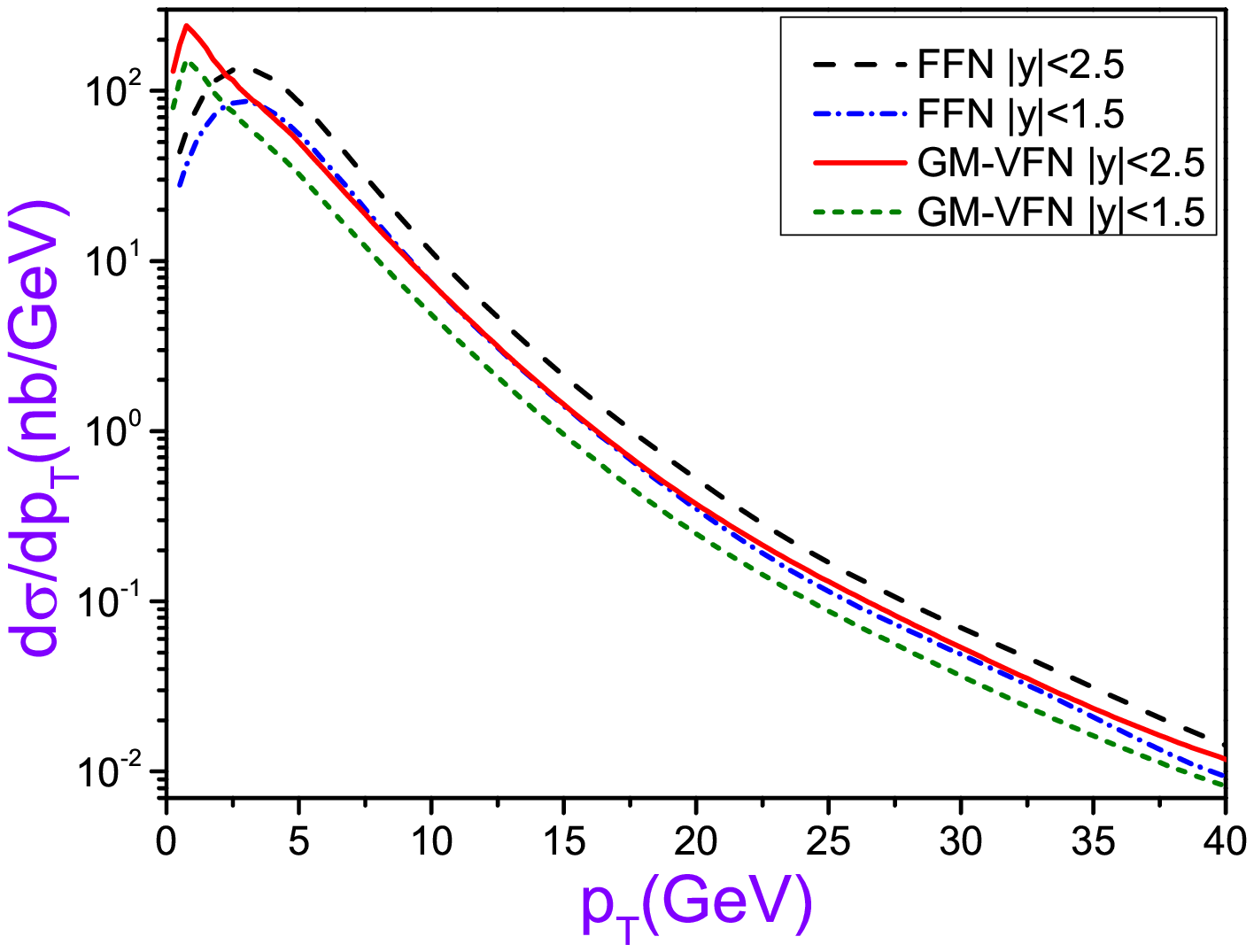}
\caption{$p_T$ distributions for the hadronic production of $B_s$ (Left) and $B_s^*$ (Right) at LHC with $\sqrt S=14.0$ TeV. The dashed and the dash-doted lines are for gluon-gluon fusion results obtained under the FFNS for rapidity cuts $|y|<2.5$ and $|y|<1.5$, respectively. The solid and the short-dashed lines stand for the total (The sum of the extrinsic $b$-quark and $s$-quark mechanism and the gluon-gluon fusion mechanism) results obtained under the GM-VFNS for rapidity cuts $|y|<2.5$ and $|y|<1.5$ respectively.} \label{fig4}
\end{figure}

\begin{figure}[ht]
\centering
\includegraphics[width=0.49\textwidth]{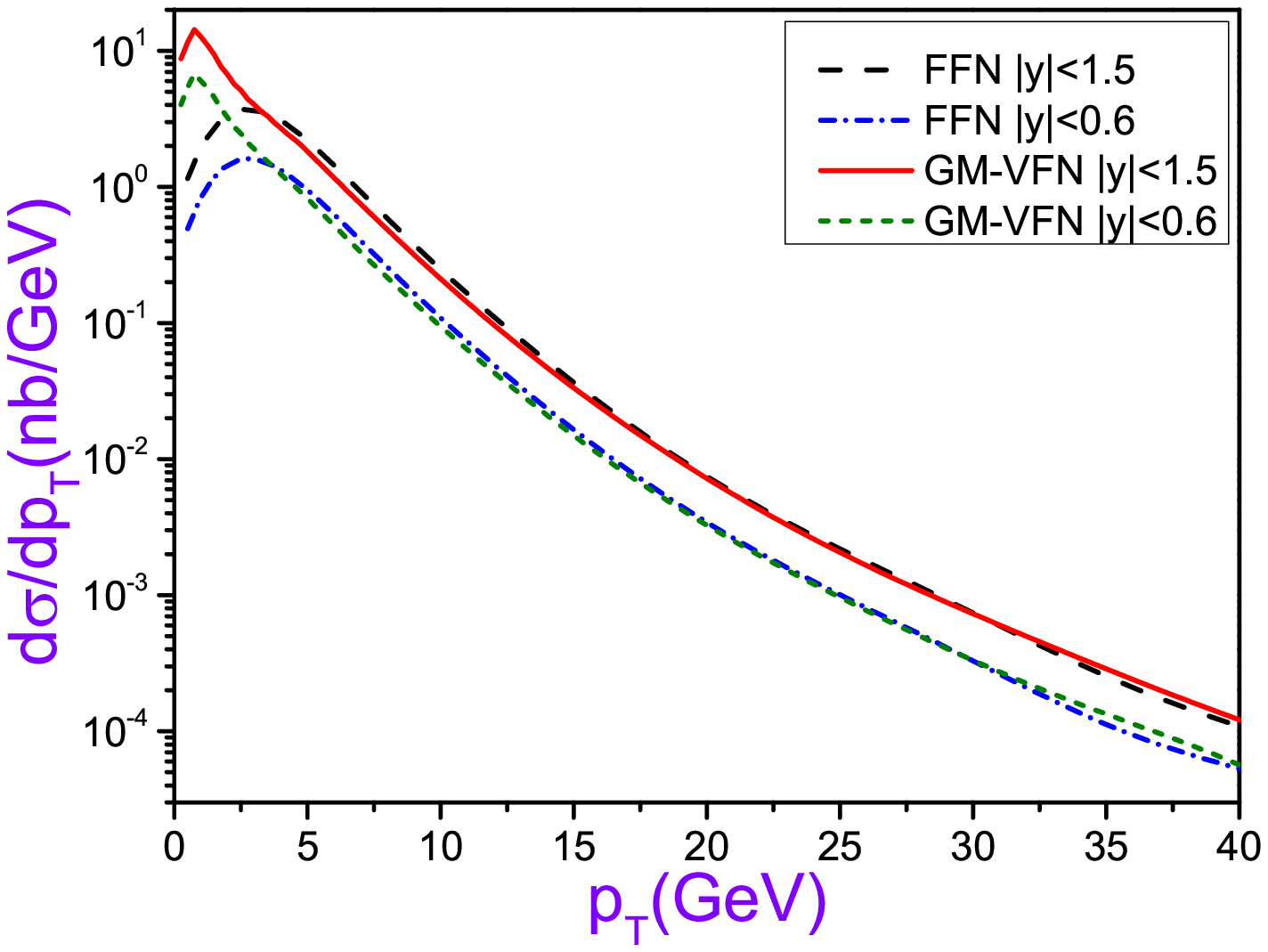}
\includegraphics[width=0.49\textwidth]{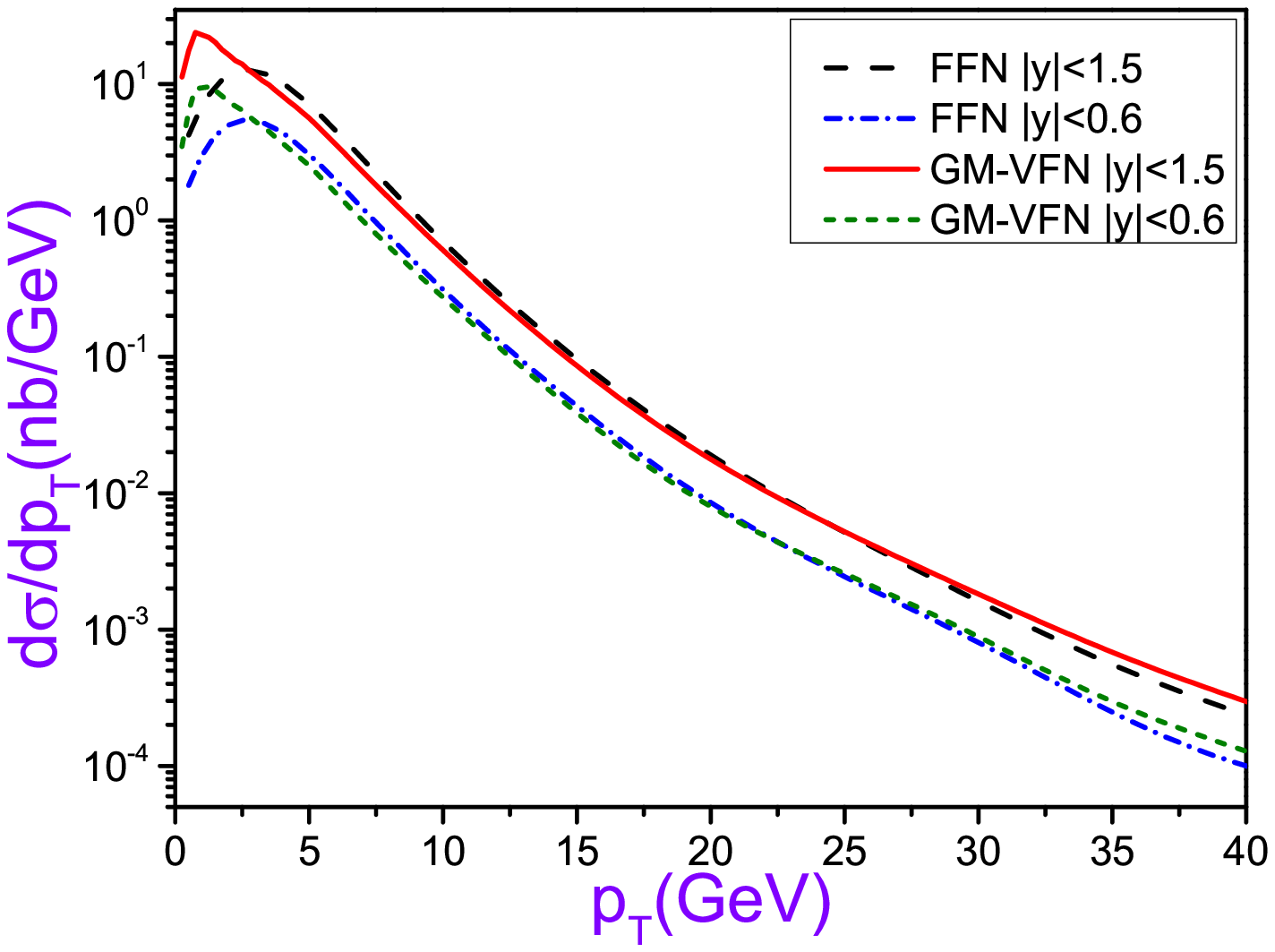}
\caption{$p_T$ distributions for the hadronic production of $B_s$ (Left) and $B_s^*$ (Right) at Tevatron. The dashed and the dash-doted lines are for gluon-gluon fusion results obtained under the FFNS for rapidity cuts $|y|<1.5$ and $|y|<0.6$, respectively. The solid and the short-dashed lines stand for the total (The sum of the extrinsic $b$-quark and $s$-quark mechanism and the gluon-gluon fusion mechanism) results obtained under the GM-VFNS for rapidity cuts $|y|<1.5$ and $|y|<0.6$ respectively. } \label{fig5}
\end{figure}

In order to see the fact clearly, we present the $p_T$ distributions predicted by the GM-VFN and FFNS for the hadronic production of $B^{(*)}_s$ at the LHC and the Tevatron in Figs.(\ref{fig3},\ref{fig4},\ref{fig5}) respectively. Figs.(\ref{fig3},\ref{fig4},\ref{fig5}) show that the main difference between the predictions by the GM-VFNS and the FFNS is only in small $p_T$ region ($p_T\lesssim 3.0\sim 4.0$ GeV). And the results under GM-VFNS and FFNS are consistent with each other in the large $p_T$ regions. Especially, Fig.(\ref{fig5}) shows that at Tevatron, one can hardly distinguish the difference between FFNS and GM-VFNS, since a $p_{Tcut}\simeq 4 GeV$ is practically adopted at Tevatron for analyzing the hadronic productions. This shows that at the Tevatron, both GM-VFNS and FFNS can describe the data consistently. While Fig.(\ref{fig3}) and (\ref{fig4}) shows that at the LHC with $\sqrt S=8.0$ TeV and $\sqrt S=14.0$ TeV, such difference is amplified, so the forthcoming LHC experiment data may make a judge on whether we need to take the heavy quark component in proton into consideration (and hence the necessity of using GM-VFNS), since more small $x$ and small $p_T$ events can be found/measured at the LHC.

\section{Summary}

We have suggested two mechanisms, e.g. the gluon-gluon fusion mechanism and the extrinsic heavy quark mechanism, for the $B^{(*)}_s$ meson hadroproduction. Under the FFNS, we only need to deal with the dominant gluon-gluon fusion mechanism~\cite{zfw}. At the present paper, we have reanalyzed it under the GM-VFNS, in which these two mechanisms should be taken into consideration so as to make a sound estimation. In our calculation, we have treated the $s$-quark as heavy quark, which is reasonable and our results show that the heavy $s$-quark approximation can lead to reasonable estimations under both GM-VFNS and FFNS.

To be useful reference, a comparison of the estimations under FFNS and GM-VFNS is presented. It is found that the extrinsic mechanism can be as important as the gluon-gluon fusion mechanism. Especially, in small $p_T$ region, the extrinsic mechanisms are even dominant over the gluon-gluon fusion mechanism, which are clearly shown in Figs.(\ref{fig1},\ref{fig2}). However, the cross section for the extrinsic mechanism drops down much quickly with the increment of $p_T$. More explicitly, total cross sections versus several typical $p_{Tcut}$ are shown in Tables \ref{cutcross1} and \ref{cutcross2}. These two tables show that by setting $p_{Tcut}=0$ GeV, the ratio between the total cross-section of the extrinsic mechanism and that of the gluon-gluon fusion mechanism are $160\%$ for $B_s$ and $80\%$ for $B^{*}_s$ at both LHC and Tevatron; and when $p_{Tcut}$ increases to 2.5 GeV and 4.0 GeV, such ratio changes down to $\sim 23\%$ and $\sim 9\%$ for the case of $B_s$, and $\sim 15\%$ and $\sim 5\%$ for the case of $B^{*}_s$, respectively.

More data to come provides us chances to know more subtle structures of the hadron, such as the extrinsic or the intrinsic heavy quark components~~\cite{hadron}. For the present case, it is noted that the extrinsic heavy quark mechanism can be used as a supplement to the usual gluon-gluon fusion mechanism. Our results show that similar to the case of the double heavy mesons / baryons as $B_c$, $\Xi_{cc}$ and etc., if the hadronic experiments such as those at LHC can accumulate large enough data and measure low $p_T$ events, then they can provide a good platform to check which scheme, either GM-VFNS or FFNS, is more appropriate for studying the heavy quark properties. Probably some suitable fixed target experiments, such as the suggested AFTER@LHC~\cite{sjb}, in which the detector may cover almost all solid angles (almost without $p_t$ cut), can test the extrinsic heavy quark mechanism in the future.

In the large $p_T$ region, the estimates of GM-VFNS and FFNS are close in shape, because in this region, GM-VFNS is also dominated by the gluon-gluon fusion mechanism. As a subtle point, it is noted that in the large $p_T$ region, we will have large logarithms of $(p_T/m_Q)$ ($m_Q$ being the heavy quark mass), which makes the pQCD convergence sometimes questionable. It is argued that one can resum all those large logs to achieve a more convergent estimation. For example, in the literature, the FONLL resummation under the fragmentation approach has been suggested for dealing with the $B$-meson production~\cite{fonll1,fonll3}. We hope a similar resummation for the present process may further shrink the gap between the GM-VFNS and FFNS in the large $p_T$ region. \\

{\bf Acknowledgments:} This work was supported in part by Research Foundation of Chongqing University of Science \& Technology under Grant No.CK2011B34, by Natural Science Foundation of China under Grant No.11075225 and No.11275280, by the Program for New Century Excellent Talents in University under Grant No.NCET-10-0882, and by the Fundamental Research Funds for the Central Universities under Grant No.WLYJSBJRCGR201106 and No.CQDXWL-2012-Z002.

\end{document}